\documentclass[jkps,preprint,fleqn,showpacs,showkeys]{revtex4}
\usepackage{graphicx}
\usepackage{amssymb}
\usepackage{amsmath}
\usepackage{bm}
\usepackage{multirow}
\usepackage{enumitem}

\newcommand{\gsim}{\gtrsim}
\newcommand{\lsim}{\lesssim}

\newcommand{\beq}{\begin{equation}}
\newcommand{\eeq}{\end{equation}}
\newcommand{\bea}{\begin{eqnarray}}
\newcommand{\eea}{\end{eqnarray}}
\newcommand{\barr}{\begin{array}}
\newcommand{\earr}{\end{array}}
\newcommand{\bc}{\begin{center}}
\newcommand{\ec}{\end{center}}
\newcommand{\bit}{\begin{itemize}}
\newcommand{\eit}{\end{itemize}}
\newcommand{\ben}{\begin{enumerate}}
\newcommand{\een}{\end{enumerate}}

\newcommand{\nn}{\nonumber}
\newcommand{\sm}{{\rm SM}}
\newcommand{\br}{{\rm B}}

\newcommand{\ifb}{{\,{\rm fb}^{-1}}}

\newcommand{\gev}{{\;{\rm GeV}}}
\newcommand{\tev}{{\;{\rm TeV}}}
\newcommand{\al}{\alpha}

\newcommand{\sg}{\sigma}
\newcommand{\gm}{\gamma}

\newcommand{\lm}{\lambda}

\newcommand{\tb}{t_\beta}

\newcommand{\cba}{c_{\beta-\alpha}}
\newcommand{\sba}{s_{\beta-\alpha}}

\newcommand{\mch}{M_{H^\pm}}

\newcommand{\hsm}{h_{\rm SM}}

\newcommand{\ee}{{e^+ e^-}}

\begin{document}
\setcounter{page}{0}
\title[]{Constraints from Heavy Higgs boson masses \\
in the two Higgs doublet model }
\author{Jin-Hwan  \surname{Cho}}
\affiliation{National Institute for Mathematical Sciences,
Yuseong-gu, Daejeon 34047, Korea}
\author{Tae Young \surname{Kim}}
\author{Jeonghyeon  \surname{Song}}
\email{jhsong@konkuk.ac.kr}
\affiliation{Department of Physics, Konkuk  University, Gwangjin-gu, Seoul
135-703, Korea}


\begin{abstract}
Upon the absence of signals of new physics at the LHC,
a reasonable strategy  is to 
assume that new particles are very heavy 
and the other model parameters are unknown yet.
In the aligned two Higgs doublet model,
however,
heavy Higgs boson masses 
above 500 GeV enhance some couplings in the scalar potential, 
which causes a breakdown of the perturbative unitariry in general.
Some tuning among model parameters is required.
We find that one information on the heavy Higgs boson mass, say $M_H$,
has significant theoretical implications:
(i) the other heavy Higgs bosons should have similar masses
to $M_H$ within $\pm \mathcal{O}(10)\%$;
(ii) the inequalities from the theoretical constraints
are practically reduced to an equation such that
$m_{12}^2 \tan\beta$ is constant, where $m_{12}^2$
is the soft $Z_2$ breaking parameter and $\tan\beta$ is
the ratio of two vacuum expectation values;
(iii) the triple Higgs coupling $\lambda_{HHh}$ is constant over $\tan\beta$
while $\lambda_{HHH}$ and $\lambda_{AAH}$
are linearly proportional to $\tan\beta$.
The double Higgs-strahlung process of $e^+ e^- \to ZHH$
is also studied, of which the total cross section is almost constant 
with the given $M_H$. 
\end{abstract}



\maketitle

\section{INTRODUCTION}
The standard model (SM) of particle physics
seems to complete the journey by the 
discovery of the last missing piece of the puzzle,
a Higgs boson with a mass of 125 GeV~\cite{Aad:2012tfa,Chatrchyan:2012xdj}.
This newly discovered scalar boson is very likely the SM Higgs boson,
according to the combined analysis by the ATLAS and CMS experiments~\cite{Khachatryan:2016vau,Aad:2015gba}:
the signal yield of $\sigma_i \cdot \br_f$ is $1.09 \pm 0.11$ of the SM prediction.
The observation of the SM-like Higgs boson
provides a basecamp for the next level questions.
A significant one is
whether the observed Higgs boson is the only fundamental scalar boson.
Many new physics models predict additional scalar bosons~\cite{Ivanov:2017dad},
which
get constraints from the Higgs precision data~\cite{Christensen:2012ei,Athron:2017qdc,Aggleton:2016tdd,Belanger:2013xza,Cheung:2013kla,Chiang:2013ixa,Chang:2012ve,Chang:2013ona,Chang:2015goa}.
In addition, the null results of
the dedicated searches for new scalar bosons at the LHC~\cite{Aaboud:2017fgj,Aaboud:2017itg,Aaboud:2017eta,Aaboud:2017yyg,Khachatryan:2016yec,Aaboud:2017sjh}
also limit the models.

The constraints become very strong when the observed
Higgs boson $\hsm$ is 
assigned as a \emph{heavier} Higgs boson state, say $H$.
Adjusting $H$ to behave almost the same as $\hsm$
and hiding the lighter Higgs boson states from low energy experiment data 
constrain a new physics model very tightly.
This feature was studied in
the two Higgs doublet model (2HDM) with \textit{CP} invariance and softly broken $Z_2$ symmetry~\cite{Bernon:2015wef}.
There exist five physical Higgs bosons,
the light \textit{CP}-even scalar $h$,
the heavy \textit{CP}-even scalar $H$, the \textit{CP}-odd pseudoscalar $A$,
and two charged Higgs bosons $H^\pm$~\cite{Gunion:2002zf,Branco:2011iw}.
The survived parameter space is meaningfully limited.
For example,
this hidden light Higgs scenario does not allow
the $A$ and $H^\pm$ heavier than about 600 GeV~\cite{Chang:2015goa}.

When we set the observed Higgs boson to be the lighter $h$,
on the contrary,
the new physics model seems to have much more freedom.
The absence of new signals of additional scalar bosons 
is explained by taking the decoupling limit where 
the other Higgs states are outside the LHC reach~\cite{Chakrabarty:2014aya, Coleppa:2017lue}.
The Higgs precision data can be easily accommodated 
by the scalar alignment limit in the 2HDM~\cite{Gunion:2002zf}.
The aligned 2HDM with decoupling~\cite{Carena:2013ooa,Dev:2014yca}
is expected to be safe, albeit unattractive.
There have been great experimental efforts to hunt for heavy neutral scalar bosons
in various channels, as summarized in Table \ref{tab:Exp}.
Most searches target the mass range up to a few TeV.
Examining the heavy scalar search strategies altogether,
we note that 
the masses and couplings of the heavy Higgs bosons
are treated to be free even in a specific model like the 2HDM.
In the viewpoint of free parameter setting,
this approach is reasonable since
the 2HDM can be described by the physical parameters of
$m_h$, $M_H$, $M_A$, $M_{H^\pm}$, $m_{12}^2$,
$\tan\beta$, and $\alpha$~\cite{Cheung:2003pw},
where $m_{12}^2$ is the soft $Z_2$ symmetry breaking term
while $\alpha$ and $\beta$ are two mixing angles
in the Higgs sector.
The big advantage of the choice is that
one measurement like a heavy Higgs boson mass
is directly related with one model parameter.
Being independent of each other as free parameters,
the other parameters require new measurements.

\begin{table}
\caption{\label{tab:Exp} The current status of the searches for heavy scalar bosons
at the LHC.}
\begin{ruledtabular}
\begin{tabular}{c|cccc}
~~channel~~ & $\sqrt{s}$ & $\int \mathcal{L}d t$ & $M_S$ range & Experiment \\
\colrule
%
 $WW/ZZ$ &13 TeV & 36 $\ifb$ & $0.3-3\tev$ & ATLAS~\cite{Aaboud:2017fgj,Aaboud:2017itg,Aaboud:2017eta}\\ \colrule
\multirow{2}{*}{ $\gamma\gamma$ }& 13 TeV & 36 $\ifb$ & $0.2-2.7\tev$ & ATLAS~\cite{Aaboud:2017yyg}\\
 & 8+13 TeV & 19.7+16.2 $\ifb$ & $0.5-4.5\tev$ & CMS~\cite{Khachatryan:2016yec} \\ \colrule
$\tau\tau$ & 3 TeV & 36.1 $\ifb$ & $0.2-2.25 \tev$ & ATLAS~\cite{Aaboud:2017sjh}
\end{tabular}
\end{ruledtabular}
\end{table}

This independence is not perfectly maintained when we consider  
another class of important constraints,
the theoretical stability of the model.
The theory should maintain the unitarity, perturbativity, a
bounded-from-below scalar potential, and the vacuum stability.
Since these constraints are expressed by inequalities,
the physical parameters have been considered still free in many studies.
However in some cases the inequalities become too difficult 
to satisfy:
only very narrow parameter space survives.
We find that this happens when non-SM Higgs bosons are very heavy and
$\tan\beta$ is large.
In this limit,
one coupling in the Higgs potential is
proportional to $(M_H \tan\beta)^2$, too large
to preserve the theoretical stability generally.
Extremely narrow parameter space survives,
which yields strong correlations among the model parameters,
especially between $\tan\beta$ and $m_{12}^2$.
Consequently the theoretical constraints
shall limit the Higgs triple couplings
and the double Higgs-strahlung process
in the future $e^+ e^-$ collider.
These are our main results.

\section{The brief review of the 2HDM}
We consider a 2HDM
with \textit{CP} invariance
where there are two complex $SU(2)_L$ Higgs doublet scalar fields:
\bea
\Phi_1 = \left( \begin{array}{c} \phi_1^+ \\
\dfrac{v_1 +  \rho_1 + i \eta_1 }{ \sqrt{2}} 
\end{array} \right),
\quad 
\Phi_2 = \left( \begin{array}{c} \phi_2^+ \\
\dfrac{v_2 +  \rho_2 + i \eta_2 }{ \sqrt{2}} 
\end{array} \right).
\eea
Here $v_1 = v \cos\beta$, $v_2 = v\sin\beta$, and $v=246\gev$ is the vacuum expectation value
(VEV) of the SM Higgs field.
For notational simplicity, we adopt $c_x=\cos x$, $s_x = \sin x$,
and $t_x= \tan x$ in what follows.
In order to avoid the tree level FCNC,
we introduce the $Z_2$ parity symmetry
under which $\Phi_1 \to \Phi_1$, $\Phi_2 \to -\Phi_2$, $Q_L \to Q_L$, and $L_L \to L_L$.
Here $Q_L$ and $L_L$ are the left-handed quark and lepton doublets, respectively.
Then each right-handed fermion field couples to only one scalar doublet field.
There are four different ways to assign the $Z_2$ symmetry on the SM fermion fields,
leading to four different types in the 2HDM, Type I, Type II, Type X (leptophilic), and Type Y.

The Higgs potential is written as 
\begin{eqnarray}
\label{eq:VH}
V_H=&&m^2_{11} \Phi^{\dagger}_1 \Phi_1 + m^2_{22} \Phi^{\dagger}_2 \Phi_2 -(m^2_{12} \Phi^{\dagger}_1\Phi_2+h.c) \nn \\
&& +\frac{1}{2} \lambda_1 (\Phi^{\dagger}_1 \Phi_1)^2+
\frac{1}{2} \lambda_2 (\Phi^{\dagger}_2 \Phi_2)^2+
\lambda_3  (\Phi^{\dagger}_1 \Phi_1)  (\Phi^{\dagger}_2 \Phi_2) +
\lambda_4  (\Phi^{\dagger}_1 \Phi_2)  (\Phi^{\dagger}_2 \Phi_1) \nn \\
&& +\frac{1}{2} \lambda_5 
\left[ (\Phi^{\dagger}_1 \Phi_2)^2+ (\Phi^{\dagger}_2 \Phi_1)^2 \right],
\end{eqnarray}
where $m_{12}^2$ is the soft $Z_2$ symmetry breaking term,
which can be negative.
The mass eigenstates of $h$, $H$, $A$, and $H^\pm$ 
are defined through two mixing angles, $\al$ and $\beta$,
as
\bea
\left(
\begin{array}{c}
h_1 \\ h_2
\end{array}
\right) =
\mathbb{R}(\al) 
\left(
\begin{array}{c}
H \\ h
\end{array}
\right),
\quad
\left(
\begin{array}{c}
\eta_1 \\ \eta_2
\end{array}
\right) =
\mathbb{R}(\beta) 
\left(
\begin{array}{c}
z \\ A
\end{array}
\right)
, \quad
\left(
\begin{array}{c}
w_1^\pm \\ w_2^\pm
\end{array}
\right) =
\mathbb{R}(\al) 
\left(
\begin{array}{c}
w^\pm \\ H^\pm
\end{array}
\right),
\eea
where $z$ and $w^\pm$ are the Goldstone bosons to be eaten by the $Z$ and $W$ bosons, respectively.
The rotation matrix $\mathbb{R}(\theta)$ is
\bea
\mathbb{R}(\theta) = \left(
\begin{array}{cc}
c_\theta & -s_\theta \\ s_\theta & c_\theta
\end{array}
\right).
\eea

In order to explain the Higgs precision data and the heavy Higgs search results,
we take the alignment limit with decoupling.
Brief comments on the terminologies of the decoupling and alignment regime
are in order here.
The decoupling regime corresponds to the parameter space where all of the extra Higgs bosons are
much heavier than the lightest Higgs boson $h$.
The terminology alignment is used in two different ways.
It was first used in the Yukawa sector to avoid the tree-level FCNC 
without introducing the $Z_2$ symmetry~\cite{Pich:2009sp}.
The second way is the scalar alignment, leading to $\hsm =h$.
Upon the observed SM-like Higgs boson, the scalar alignment is 
commonly abbreviated as the alignment, which is adopted here. 
Since $\hsm = \sba h + \cba H$,
the alignment requires 
\bea
\label{eq:alignment}
\hbox{alignment limit: } \sba =1.
\eea
We note that in the alignment limit
the following couplings among $V_\mu (= Z_\mu, W^\pm_\mu)$ and the heavy 
Higgs bosons vanish:
\bea
\label{eq:0couplings}
\sba=1:~Hhh, \quad V_\mu V_\nu H, \quad Z_\mu h A, \quad V_\mu V_\nu H h 
\quad {\longrightarrow} ~~0 
.
\eea

The potential $V_H$ has eight parameters of
$ m_{11}^2$, $m_{22}^2$, $m_{12}^2$, $\lm_1$, $\lm_2$, $\lm_3$, $\lm_4$,
and $\lm_5$.
Two tadpole conditions replace $ m_{11}^2$ and $m_{22}^2$
by the known $v$ and the unknown $\tb$ through
\bea
\label{eq:m11sq:m22sq}
m_{11}^2 &=& m_{12}^2 \tb - \frac{\lm_1 + \lm_{345} \tb^2}{2 (1+\tb^2)} v^2,
\\ \nn
m_{22}^2 &=& \frac{m_{12}^2 }{\tb} - \frac{\lm_2 \tb^2 + \lm_{345} }{2 (1+\tb^2)} v^2,
\eea
where $\lm_{345}=\lm_3+\lm_4+\lm_5$.
Now we have seven parameters.
Equivalently we can take the physical parameters of 
$m_h$, $M_H$, $M_A$, $M_{H^\pm}$,  $m_{12}^2$,
$\al$ and $\beta$, on which
many analyses of the aligned 2HDM are based.

In the physical parameter basis, $\lambda_i$'s $(i=1,...5)$ 
are~\cite{Kanemura:2011sj}
\begin{eqnarray}
\lambda_1 &=& \frac{1}{v^2}
\left[M^2_H \tb^2 + m^2_h - m^2_{12} \tb(1+\tb^2)
\right], \nn \\
\lambda_2 &=& \frac{1}{v^2}
\left[\frac{M^2_H}{\tb^2} + m_h^2 - m^2_{12} \frac{1+\tb^2}{\tb^3}
\right] , \nn \\
\lambda_3 &=& 
\frac{1}{v^2}
\left[
m^2_h - M^2_H + 2M^2_{H^{\pm}} - m^2_{12} \frac{1+\tb^2}{\tb}
\right] , \nn \\
\lambda_4 &=& \frac{1}{v^2}
\left[
M^2_A-2M^2_{H^{\pm}} + m^2_{12} \frac{1+\tb^2}{\tb}
\right], \nn \\
\lambda_5 &=& \frac{1}{v^2}
\left[
m^2_{12} \frac{1+\tb^2}{\tb} - M^2_A
\right]
.
\end{eqnarray}
In addition, the triple couplings of the neutral Higgs bosons 
in units of $\lm_0 = m_Z^2/v$
are written as
\bea
\label{eq:triple}
\lm_{hhh} &=& \lm_{hhh}^\sm =  \frac{ 3 m_h^2}{m_Z^2}, \quad \lm_{Hhh} =0,
\\[5pt] \nn
\lm_{HHh} &=& \frac{2 M_H^2}{m_Z^2}
\left[
1 + \frac{m_h^2}{2 M_H^2} -\frac{m_{12}^2}{2 M_H^2} \cdot \tb  \left( 1+\frac{1}{\tb^2}\right)
\right],
\\[5pt] \nn
\lm_{AAh} &=& \frac{2 M_A^2}{m_Z^2}
\left[
1+ \frac{m_h^2}{2 M_A^2} -\frac{m_{12}^2}{2 M_A^2} \cdot \tb  \left( 1+\frac{1}{\tb^2}\right)
\right],
\\[5pt] \nn 
\lm_{HHH} &=& \frac{3 M_H^2 }{m_Z^2}
\left[
1
-  \frac{m_{12}^2}{2 M_H^2} \cdot \tb^2  \left( 1-\frac{1}{\tb^4}\right)
\right],
\\[5pt] \nn
\lm_{AAH} &=&\frac{2 M_H^2}{m_Z^2}
\left[
\frac{1}{2} \left(\tb - \frac{1}{\tb}\right)
-\frac{m_{12}^2}{4 M_H^2} \cdot \tb^2  \left( 1-\frac{1}{\tb^4}\right)
\right].
\eea
Since the physical parameters are assumed free,
the triple couplings
can be any value, except for $\lm_{hhh}$ and $\lm_{Hhh}$.

For very heavy Higgs boson masses, however, this approach may lead to a breakdown of perturbative unitarity.
If $M_{H}\simeq M_A \simeq \mch\gg m_h$ and $\tb \gg 1$,
we have
\begin{eqnarray}
\label{eq:lm:limit}
\lambda_1 \simeq  \frac{\tb^2}{v^2} \left[{M_H^2} - m_{12}^2 \tb\right], \quad
\lambda_2 \simeq \frac{m_h^2}{v^2}, \quad
\lambda_3 \simeq -\lm_4\simeq -\lm_5 \simeq  \frac{1}{v^2} \left[{M_H^2} - m_{12}^2 \tb\right]
,
\end{eqnarray}
which become too large to preserve the perturbative unitarity.
We need fine tuning among model parameters.

\section{Phenomenological and theoretical constraints}
We first consider the following direct and indirect experiments involving scalar bosons:
\begin{enumerate}[label={\ttfamily \Alph*.}]
\item the LHC Higgs signal strength measurements~\cite{Khachatryan:2016vau,Aad:2015gba}:
\item the absence of new scalar boson signals at high energy colliders;
	\begin{enumerate}[label=(\roman*)]
	\item the LEP bounds on $\mch$~\cite{Patrignani:2016xqp};
	\item the Tevatron bounds on the top quark decay of $t \to H^+ b$~\cite{Abulencia:2005jd};
	\item the LHC searches for $H^\pm$~\cite{Ohman:2016gqs} and $A$~\cite{Aad:2014ioa,CMS:2014onr,Aad:2014vgg,Khachatryan:2014wca};
	\end{enumerate}
\item the indirect experimental constraints:
	\begin{enumerate}[label=(\roman*)]
	\item $\Delta\rho$ in the electroweak precision data~\cite{Higgs:Hunters:Guide,Chankowski:1999ta,Patrignani:2016xqp};
	\item the flavor changing neutral current (FCNC) data such as
	$\Delta M_{B_d}$ and $b\to s \gm$~\cite{Barger:1989fj,Bertolini:1990if,Mahmoudi:2009zx,Misiak:2017bgg}.
	\end{enumerate}
\end{enumerate}

The Higgs precision data {\ttfamily A} is 
well explained in the alignment limit.
The lighter $h$ behaves exactly the same as the SM Higgs boson.
Null results in the search for new scalar bosons ({\ttfamily B})
are explained by assuming that non-SM Higgs bosons are heavy enough.
The indirect constraints ({\ttfamily C}) require more specific conditions.
First the improved value of $\Delta\rho= 0.00040\pm 0.00024$ 
by the Higgs observation~\cite{Patrignani:2016xqp}
strongly prefers that at least two masses among $M_H$, $M_A$, and $\mch$ are degenerate~\cite{Song:2014lua}.
The FCNC processes constrain the masses of $H^\pm$ and the value of $\tb$,
which is quire strong in Type II but relaxed in Type I.
The updated next-to-next-to-leading-order SM prediction 
of $\br_\sm(\bar{B} \to X_s \gm)$~\cite{Misiak:2017bgg}
and the recent Bell result~\cite{Belle:2016ufb}
strongly bounds $\mch$ in the Type II:
$\mch > 570 ~(440) \gev$ for $\tb\gsim 2$ at 95\% (99\%) C.L.
If $\tb\lsim 2$, the $\mch$ bound rises up significantly.
Considering all of the above phenomenological constraints, 
we take the following scenario:
\bea
\label{eq:setup}
\sba=1,\quad M_{H, A, H^\pm}\geq 500\gev, \quad 2 \leq \tb \leq 40, \quad
M_i = M_j  ~(i,j=H,A,H^\pm).
\eea
Note that the theoretical implication of the \emph{heavy} scalar bosons
does not critically depend on the types of the 2HDM.

Now we impose the following theoretical constraints.
\begin{enumerate}
\item The scalar potential in Eq.~(\ref{eq:VH}) is bounded from below,
which requires~\cite{Ferreira:2014sld,Ivanov:2006yq}
\bea
\lambda_1 > 0, \quad 
\lambda_2 > 0, \quad \lambda_3 > -\sqrt{\lambda_1 \lambda_2}, \quad  
\lambda_3 + \lambda_4 - |\lambda_5| > - \sqrt{\lambda_1 \lambda_2}.  \label{eq:stability} 
\eea
\item The perturbative unitarity demands that
the following quantities are less than $8\pi$~\cite{Arhrib:2000is,Branco:2011iw}:
\bea
\label{eq:unitarity}
a_\pm & = & \frac{3}{2} (\lambda_1+\lambda_2)\pm  \sqrt{\frac{9}{4}(\lambda_1-\lambda_2)^2+(2\lambda_3+\lambda_4)^2},  \\
b_\pm & = &\frac{1}{2}\left ( \lambda_1+\lambda_2 \pm \sqrt{(\lambda_1-\lambda_2)^2+4\lambda_4^2} \right ), \nn \\
c_\pm & = & \frac{1}{2}\left ( \lambda_1+\lambda_2 \pm \sqrt{(\lambda_1-\lambda_2)^2+4\lambda_5^2} \right ), \nn \\
f_+ &=&  \lambda_3 +2\lambda_4+3\lambda_5, \quad f_- = \lambda_3 +\lambda_5, \quad f_{1}=f_2= \lambda_3 +\lambda_4, \nn \\
\nn
e_{1}&=& \lambda_3+2\lambda_4-3\lambda_5, \quad e_{2}= \lambda_3-\lambda_5, \quad p_1 = \lambda_3-\lambda_4.
\eea
\item The perturbativity requires
\bea
\label{eq:perturbativity}
\left| \lm_{i} \right| < 4 \pi, \quad i=1,\cdots,5.
\eea
\item We require that 
the vacuum of the scalar potential be global, which happens if and only if~\cite{Barroso:2013awa}
\bea
D = m_{12}^2 \left( m_{11}^2 - k^2 m_{22}^2 \right) (\tb - k) >0,
\eea
where $k=(\lm_1/\lm_2)^{1/4}$.
In practice, the vacuum stability condition 
is naturally satisfied when $M_{H,A,H^\pm} \gg m_h$ and $\tb \gg 1$:
$D \sim (m_{12}^2)^2 \tb^2$
in this limit.
\end{enumerate}

\begin{table}
\caption{\label{tab:degeneracy:conditions}
Six different cases for the required mass degeneracy with the fixed $M_H$ or $M_A$.
}
\begin{ruledtabular}
\begin{tabular}{|c|l|c||c|c|c|}
\multicolumn{3}{|c||}{fixed $M_H$} & \multicolumn{3}{c|}{fixed $M_A$}\\ 
\hline
 & \hspace{-50pt} degeneracy~~ & \hspace{-80pt} varying &
   & \hspace{-50pt}degeneracy~~ & \hspace{-80pt} varying \\ \hline
~~case-1~~ & \hspace{-50pt}$M_A =\mch$ & \hspace{-60pt}$M_A =\mch$~~ & 
\hspace{-50pt}case-4~~ & \hspace{-50pt}$M_H=\mch$~~ & \hspace{-60pt}$M_H=\mch$ \\
case-2 & \hspace{-50pt}$M_H =M_A$ & \hspace{-80pt}$\mch$ & 
\hspace{-50pt}case-5~~ & \hspace{-50pt}$M_A=M_H$~~ & \hspace{-80pt}$\mch$ \\
case-3 & \hspace{-50pt}$M_H =\mch$ & \hspace{-80pt}$M_A $ & 
\hspace{-50pt}case-6~~ & \hspace{-50pt}$M_A=\mch$~~ & \hspace{-80pt}$M_H$ \\
\end{tabular}
\end{ruledtabular}
\label{table1}
\end{table}

\begin{figure*}[t!]
\centering
\includegraphics[width=0.45\textwidth]{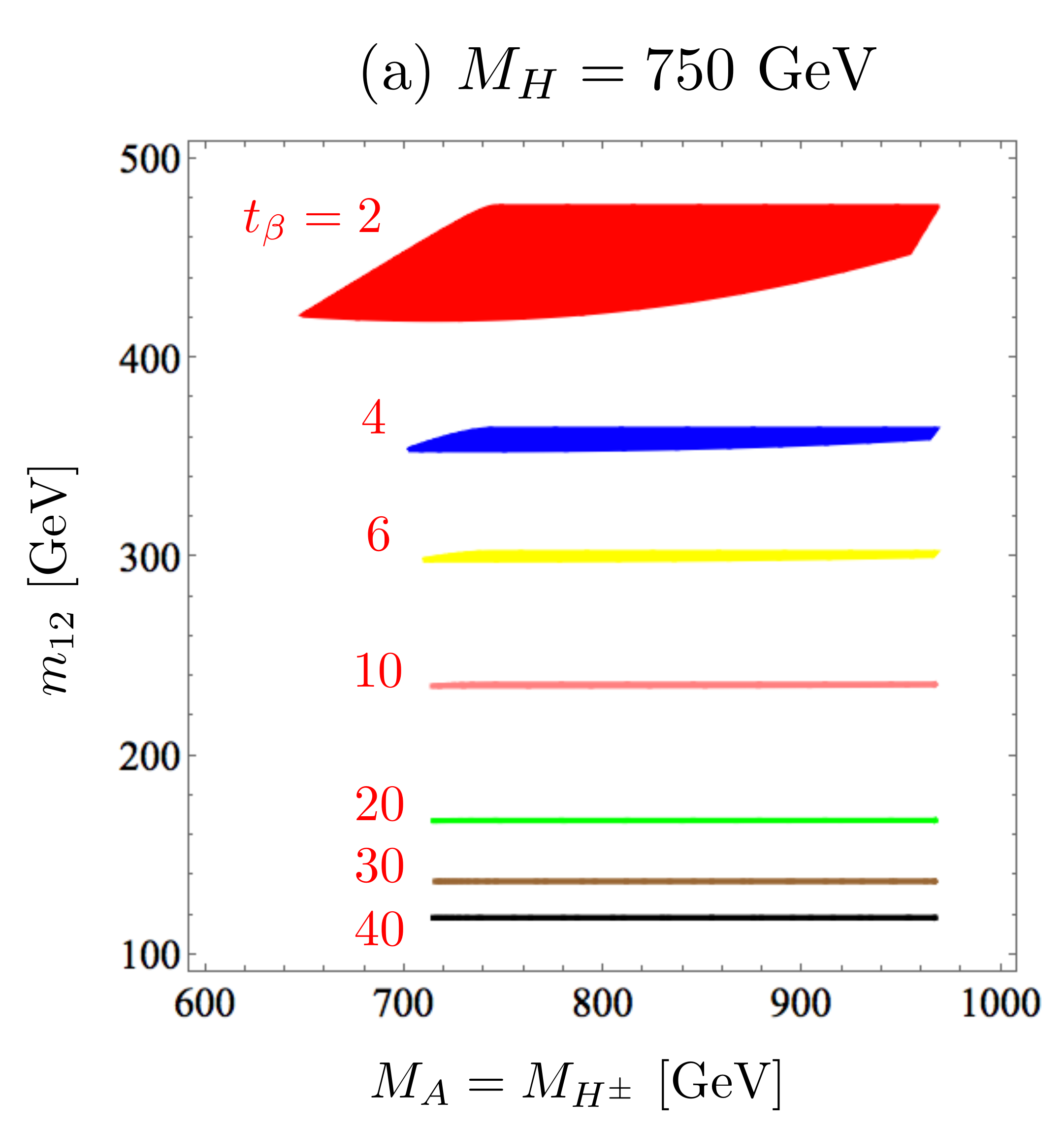}
~~
\includegraphics[width=0.45\textwidth]{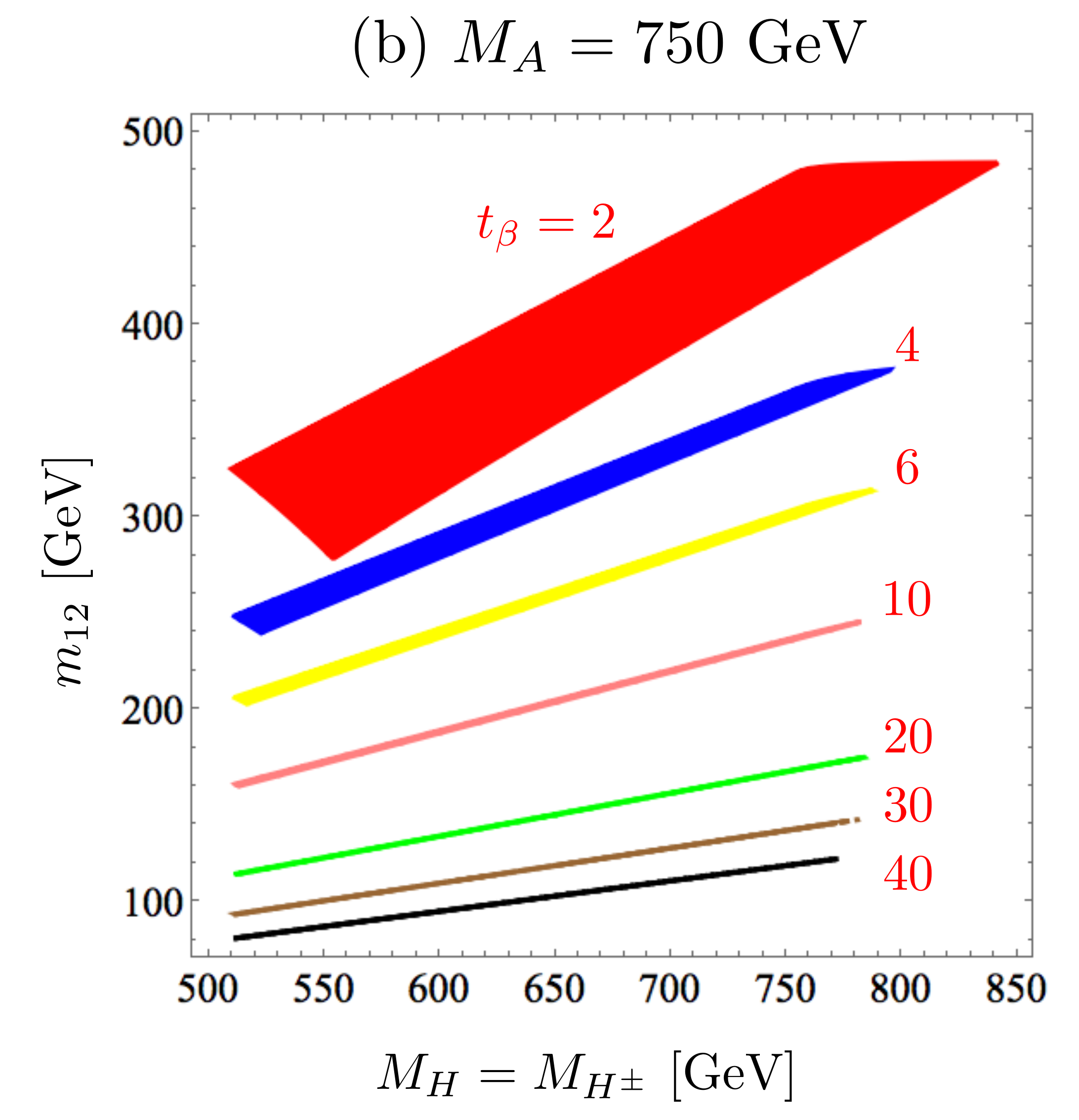}
\caption{\label{fig-M750-fix-tb}
The theoretically allowed parameter regions of $(M_{\mathcal S}, m_{12})$
for (a) $M_H=750$ GeV amd (b) $M_A=750$ GeV.
}
\end{figure*}

Now we investigate how strong the theoretical constraints are when the non-SM Higgs bosons are very heavy.
With the given $M_H$  or $M_A$, the degenerate condition in Eq.~(\ref{eq:setup})
for the $\Delta \rho$ constraint
allows six different cases as denoted in Table \ref{tab:degeneracy:conditions}.
In the case-1, for example, $M_H$ is fixed
and $M_A$ varies freely while 
$M_A=\mch$. 
In Fig.~\ref{fig-M750-fix-tb},
we show the theoretically allowed  region of $(M_S, m_{12})$ 
for the fixed $M_{H,A}$ and $\tb$.
Here $M_S$ is the other non-SM Higgs boson mass which varies freely.
Note that $m_{12}$, not $m_{12}^2$, is
presented since only positive $m_{12}^2$ is allowed by the theoretical constraints. 
We found that case-1, case-2, case-3, and case-6 show very similar allowed parameter space,
while case-4 and case-5 share almost the same allowed region.  
Therefore we present case-1 and case-4
in Fig.~\ref{fig-M750-fix-tb}(a) and (b)
as representatives.

We find some interesting results of imposing the theoretical constraints.
First the
information of one heavy Higgs boson
has significant implications on the other non-SM Higgs boson masses.
If $M_H$ is somehow measured, 
$M_{A}$ and $M_{ H^\pm}$ cannot remain as totally free parameters:
$M_H=750\gev$ requires $M_{A,H^\pm}\in[650,970]\gev$. 
Secondly there is strong correlation between $m_{12}$ and $\tb$
especially in the large $\tb$ limit.
In the fixed heavy $M_H$ case of Fig.~\ref{fig-M750-fix-tb}(a),
large $\tb$ almost fixes the value of $m_{12}$, and the dependence on $M_S$ is very weak.
The case-4 and case-5 in Fig.~\ref{fig-M750-fix-tb}(b)
also show some correlation between $m_{12}$ and $\tb$,
but weaker than in Fig.~\ref{fig-M750-fix-tb}(a):
irrespective to $M_S$, $\tb$ determines $m_{12}$ 
within the uncertainty of $\mathcal{O}(10)\gev$.
Thirdly the larger $\tb$ is, the smaller $m_{12}$ is.
Large $\tb$ prefers \emph{soft} breaking of $Z_2$ symmetry.

\begin{figure*}[t!]
\centering
\includegraphics[width=0.45\textwidth]{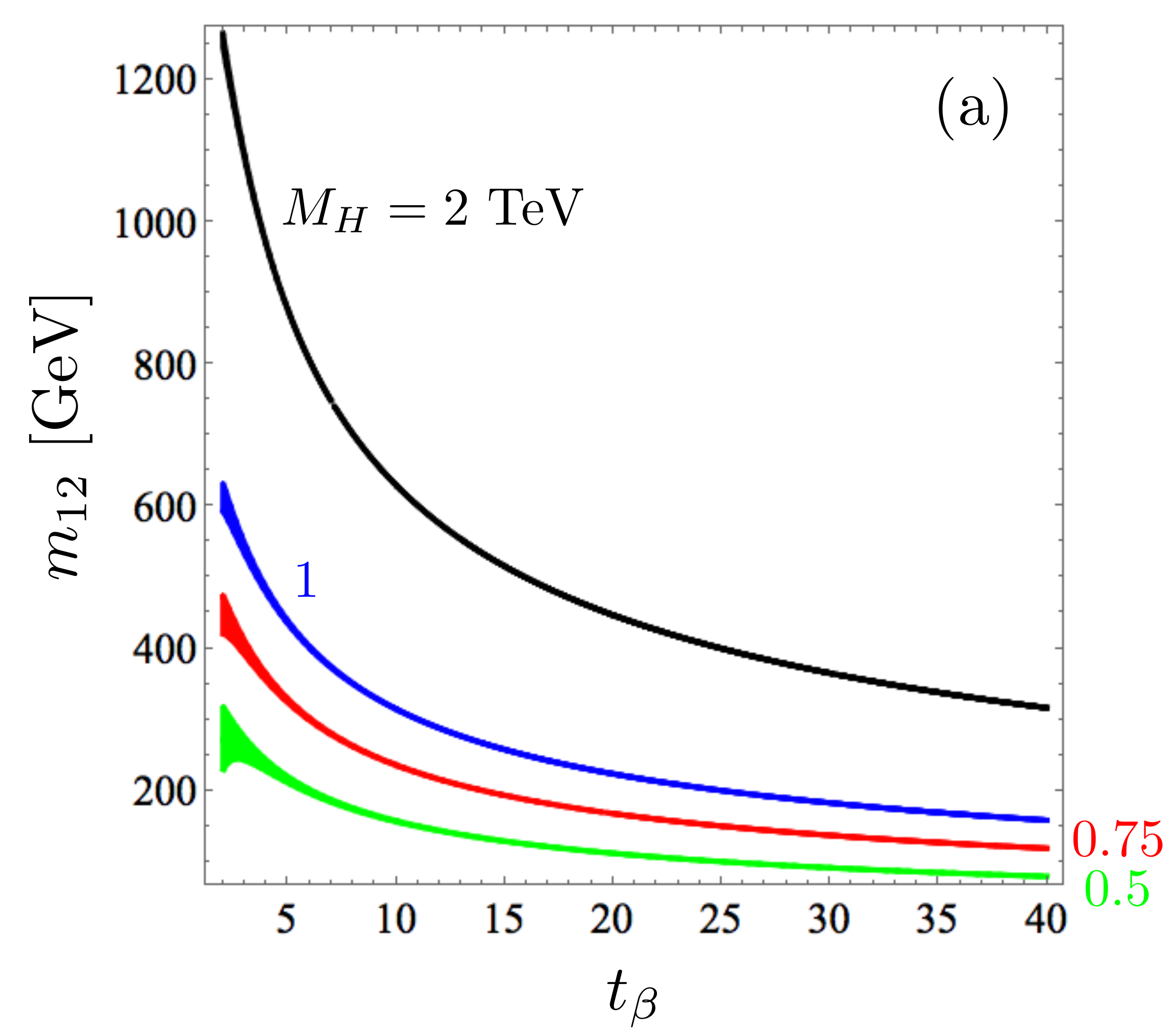}
~~
\includegraphics[width=0.45\textwidth]{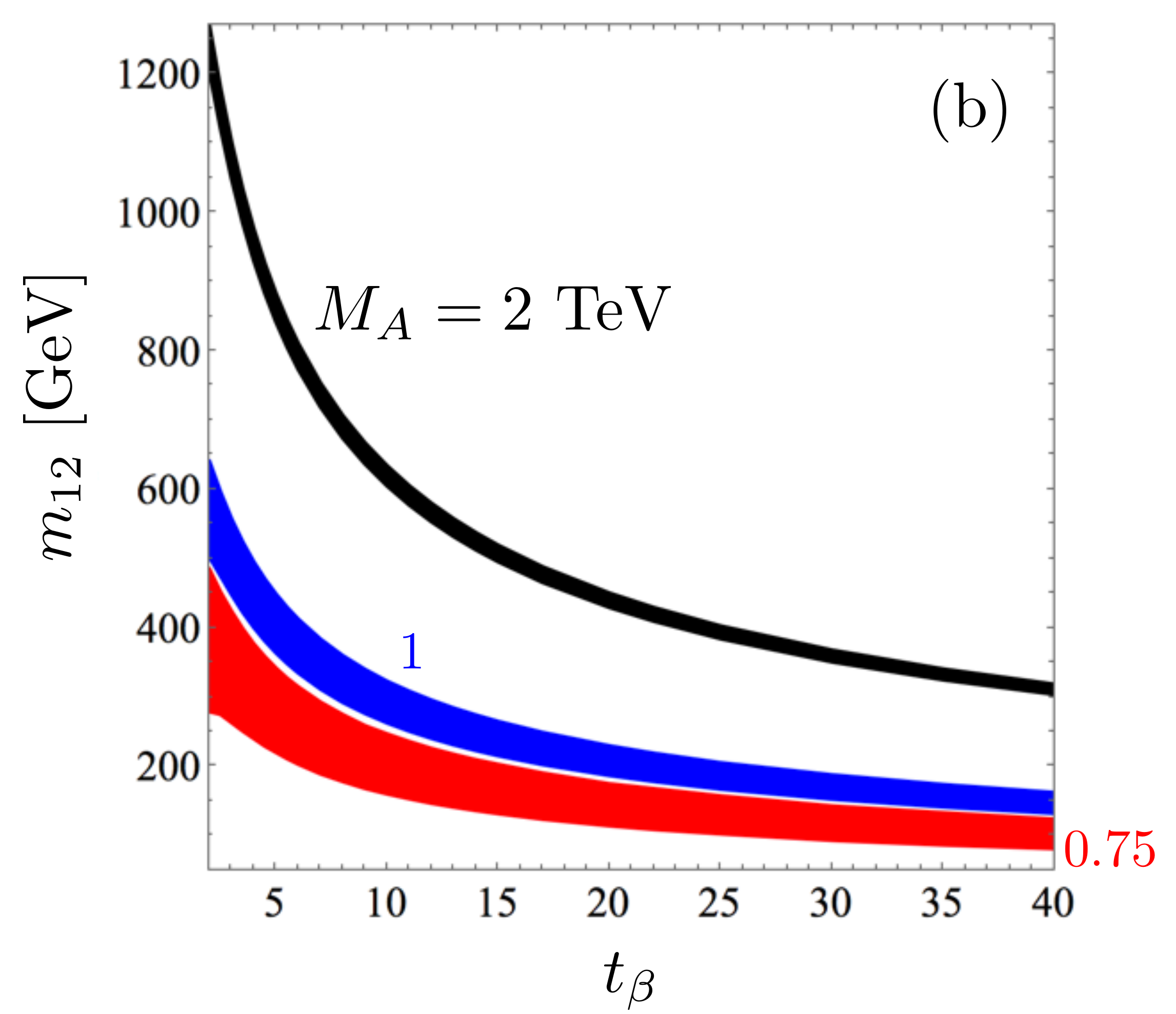}
\caption{\label{fig-M-m12-tb}
The theoretically allowed parameter regions of $(\tb, m_{12})$
for (a) $M_H=0.5$, 0.75, 1, 2 TeV and (b) $M_A=0.75$, 1, 2 TeV.
The other heavy Higgs masses vary freely.
}
\end{figure*}

In order to show the correlation between $\tb$ and $m_{12}$ more concretely,
we show the theoretically allowed region of $(\tb,m_{12})$
with the given $M_H$ or $M_A$ 
but varying $M_S$ in Fig.~\ref{fig-M-m12-tb}.
For the fixed $M_H$,
the theoretical constraints result in very strong correlation
between $\tb$ and $m_{12}$.
We find that 
the product $m_{12}^2 \tb$ becomes almost constant
especially in the large $\tb$ limit.
When $\tb \gsim 5$,
$m_{12}$ is determined within $\pm 1\gev$ by the $\tb$ value,
irrespective to $M_S$.
For the fixed $M_A$,
the correlation 
is weaker than in the fixed $M_H$ case.
The $\tb$ value sets $m_{12}$ within $\pm 10\gev$.

\begin{figure*}[t!]
\centering
\includegraphics[height=6.5cm]{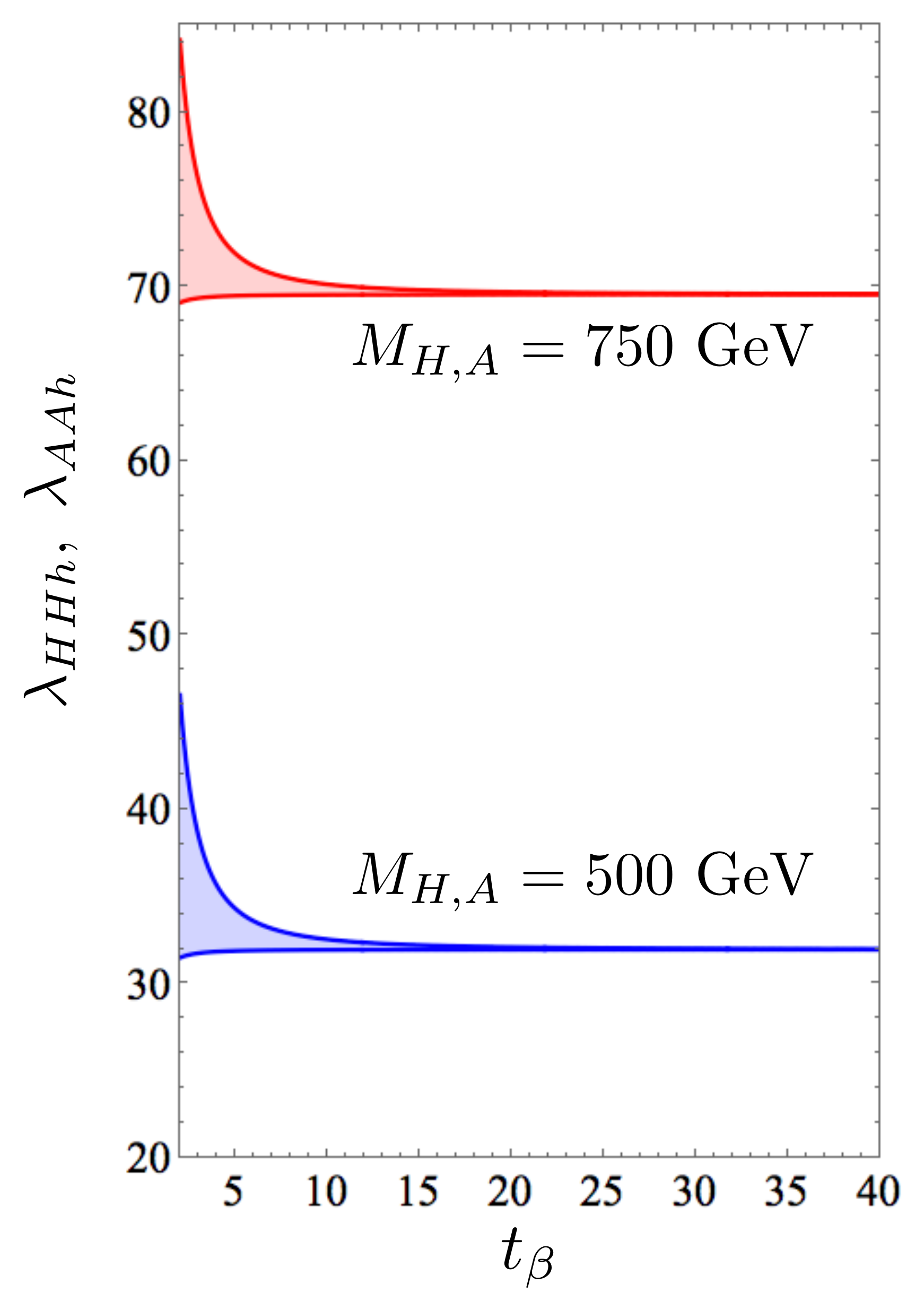}
\includegraphics[height=6.5cm]{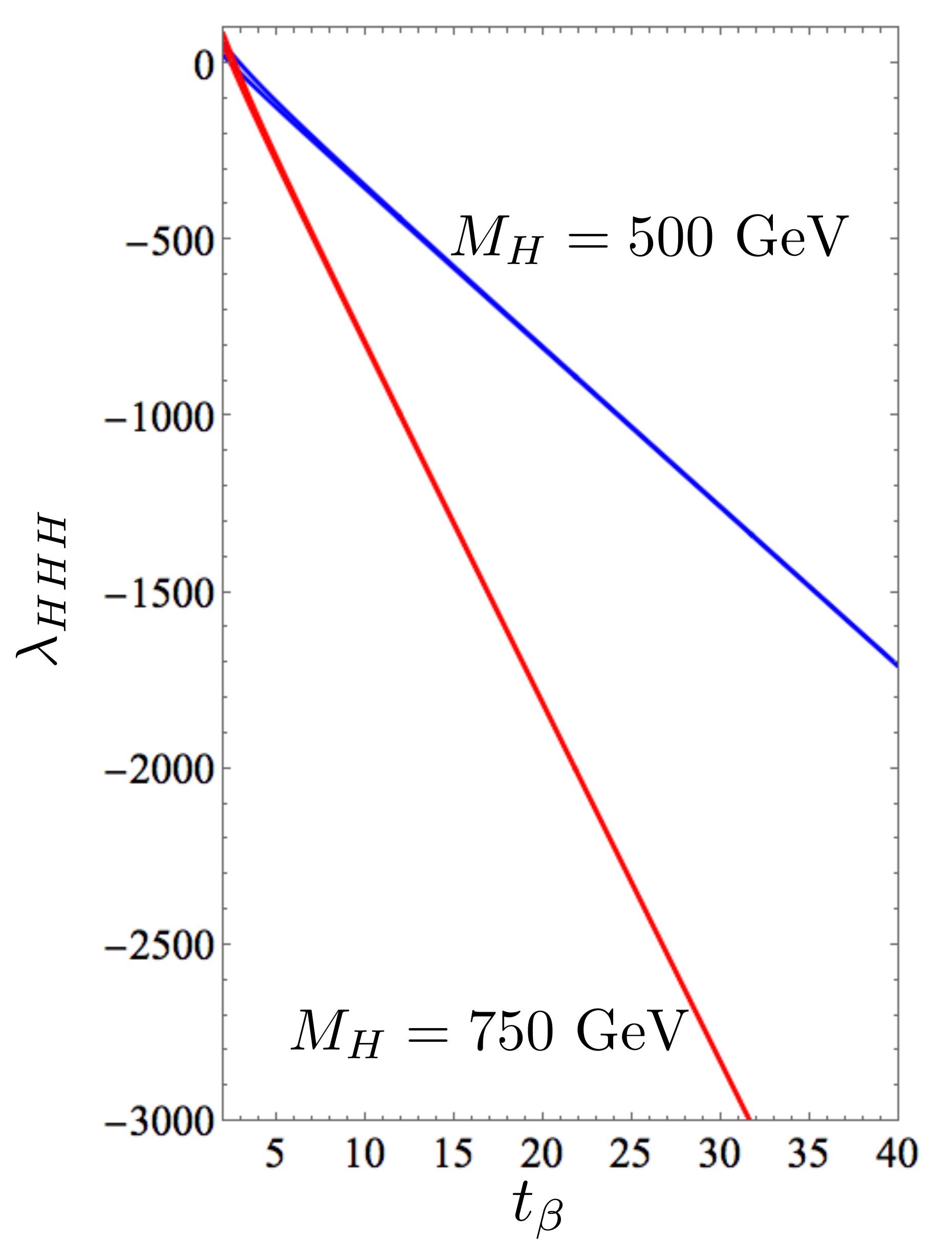}
\includegraphics[height=6.5cm]{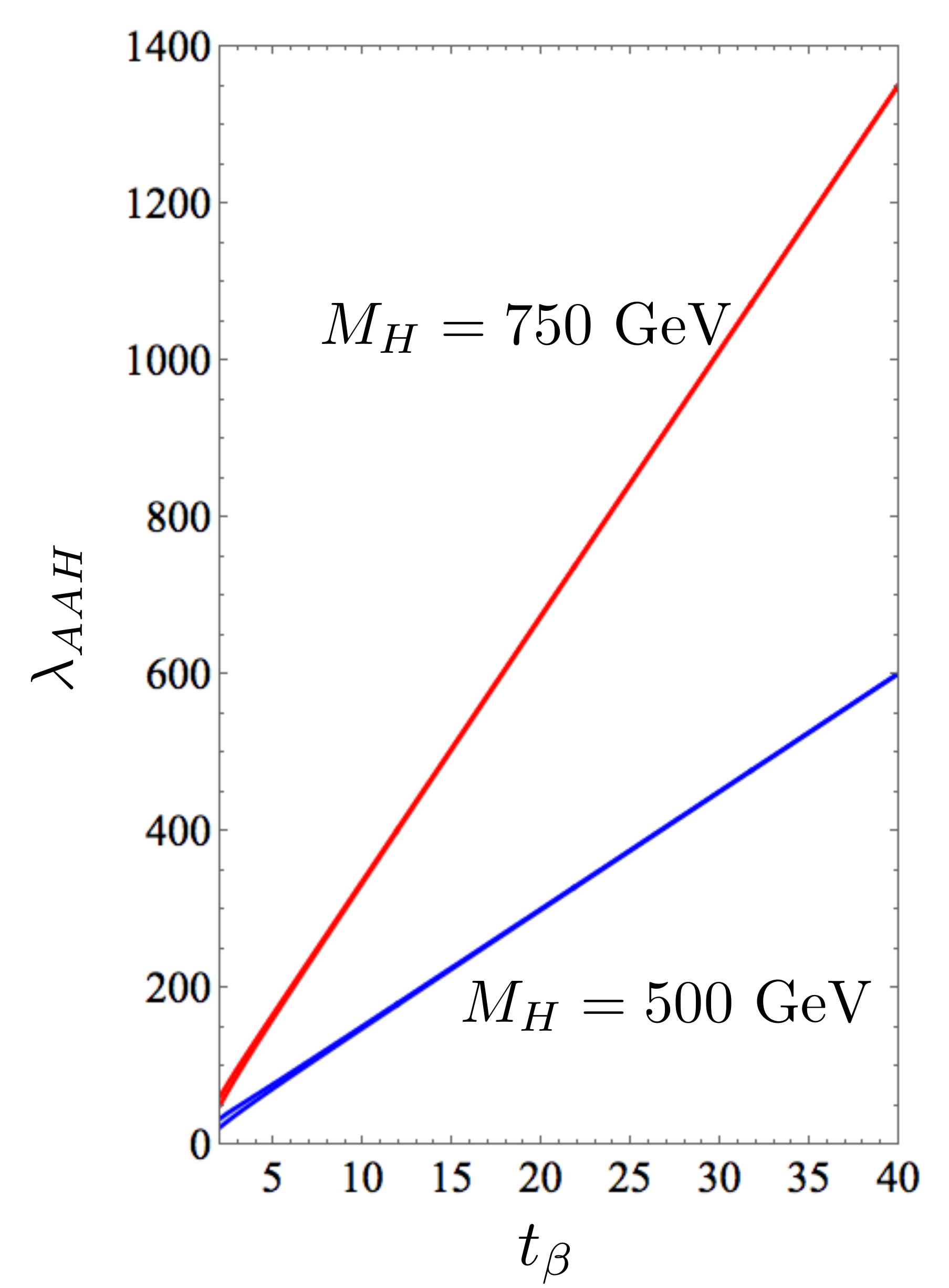}
\caption{\label{fig-lm}
The triple Higgs couplings in units of $\lm_0 =m_Z^2/v$,
allowed by the theoretical constraints for $M_{H/A} =500,750\gev$.
The other Higgs boson masses vary freely.
}
\end{figure*}

Now we study the 
theoretical constraints on the triple Higgs boson couplings.
In Fig.~\ref{fig-lm}, we present the triple Higgs couplings of
$\lm_{HHh}$, $\lm_{HHH}$, and $\lm_{AAH}$
as functions of $\tb$
in units of $\lm_0 = m_Z^2/v$
for $M_H=M_A=500,~750\gev$.
Note that when $M_H=M_A$, $\lm_{HHh}=\lm_{AAh}$: see Eq.~(\ref{eq:triple}).
The strong correlation between $\tb$ and $m_{12}$
determines the value of $m_{12}$ for the given $\tb$, and thus
the triple couplings:
effectively one model parameter $\tb$ 
almost sets the triple Higgs coupling with the given $M_H$.
As shown in Eq.~(\ref{eq:triple}),
all of the triple Higgs couplings are proportional to $M_{H,A}^2$,
which explains larger triple couplings for $M_{H,A}=750\gev$ 
than those for $M_{H,A}=500\gev$.
Interesting is their dependence on $\tb$.
The values of $\lm_{HHh}$ and $\lm_{AAh}$ becomes almost constant especially for large $\tb$,
while $\lm_{HHH}$ and $\lm_{AAH}$ are linearly proportional to $\tb$.
These behaviors are understood by the almost constant $m_{12}^2 \tb $.
In the large $\tb$ limit,
$\lm_{HHh}$
is proportional to $m_{12}^2 \tb$,
which leads to the constant behavior of $\lm_{HHh}$ over $\tb$.
Both $\lm_{HHH}$ and $\lm_{AAH}$ have dominant terms proportional to $m_{12}^2 \tb^2 $, 
which becomes linearly proportional to $\tb$ after applying the constancy
of $m_{12}^2 \tb$.

In order to probe the non-SM Higgs triple couplings,
we study 
the double Higgs-strahlung
in the future electron-positron collider~\cite{Djouadi:1999gv,Arhrib:2008jp}: 
the Future Circular Collider (FCC-ee, formally called the TLEP)~\cite{dEnterria:2016sca},
the International Linear Collider (ILC)~\cite{Baer:2013cma}, 
 the Circular Electron Positron Collider (CEPC)~\cite{CEPC-SPPCStudyGroup:2015csa},
 and the Compact Linear Collider (CLIC) \cite{Dannheim:2012rn}.
 The main target centre-of-mass (c.m.) energy of the FCC-ee and the CEPC
 is $\sqrt{s} = 240-250$ GeV,
which is suitable for the Higgs precision
measurements. 
The ILC had 
originally proposed c.m.~energy of $\sqrt{s}= 500,~800,~1000\gev$,
but the absence of new particles at the LHC and the cost reduction
lead to the decision of 250 GeV c.m.~energy~\cite{Evans:2017rvt,Asai:2017pwp}.
For the double Higgs-strahlung process involving heavy Higgs bosons, 
we need a higher c.m.~energy than 250 GeV,
which the CLIC can realize.
The CLIC is a TeV-scale linear $e^+ e^-$ 
collider, based on a two-beam technique which can
accelerate fields as high as 100 MV/m.
There are two example scenarios of energy staging:
in Scenario A (B), the c.m.~energy is 500 GeV, 1.4 (1.5) TeV and 3 TeV.
We take the case of $\sqrt{s}=1.5\tev$ with the integrated luminosity of
$1.5$ ab$^{-1}$.

In a \textit{CP} invariant framework,
there are four different double Higgs-strahlung processes at $e^+ e^-$ colliders,
$Zhh$, $ZhH$, $ZHH$, and $ZAA$.
The process of $e^+ e^- \to Zhh$ 
depends on $\lm_{hhh}$ and $\lm_{Hhh}$ in general.
In the alignment limit, however,
$\lm_{Hhh}=0$ and $\lm_{hhh}=\lm_{hhh}^\sm$: see Eq.~(\ref{eq:0couplings}).
The $Zhh$ process is the same as in the SM.
The $ZHh$ process has zero cross section at tree level
since the $ZZH$, $ZhA$, and $ZZHh$ vertices vanish in the alignment limit.
Non-vanishing and non-SM double Higgs-strahlung processes
are only two, $ZHH$ and $ZAA$, which
depend on the triple Higgs couplings as follows:
\bea
\sg(e^+ e^-\to Z H H) \ni \lm_{HHh},
\quad 
\sg(e^+ e^-\to Z A A) \ni \lm_{AAh}.
\eea 
Since $\lm_{HHh}=\lm_{AAh}$ when $M_H=M_A$,
we only consider the process of $\ee\to ZHH$.

\begin{figure*}[t!]
\centering
\includegraphics[width=0.6\textwidth]{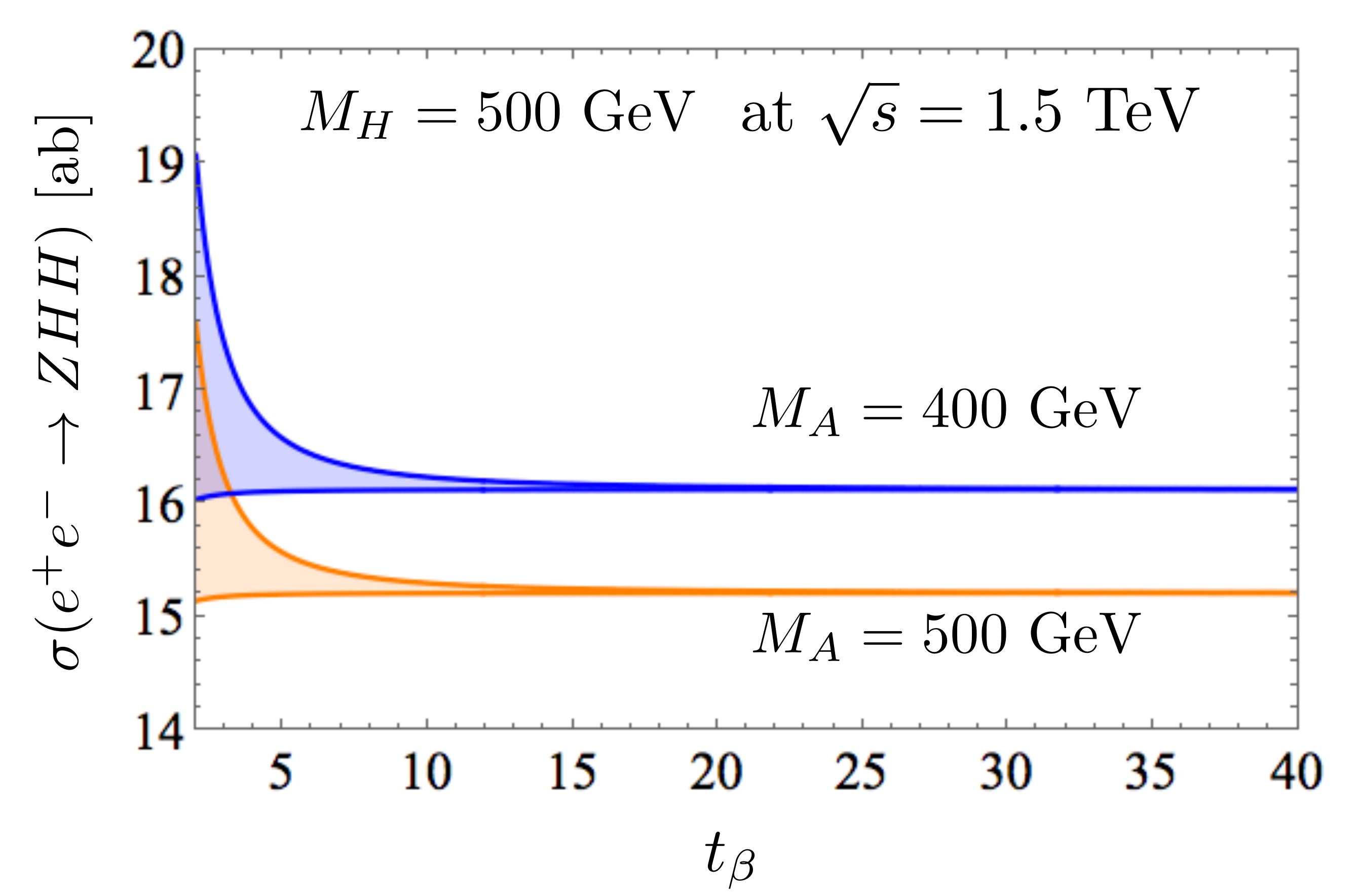}
\caption{\label{fig-sig}
The total cross section $\sg(\ee\to ZHH)$
about $\tb$,
allowed by the theoretical constraints.
We set $M_H=500\gev$, $M_A=400,500\gev$, and $\sqrt{s}=1.5\tev$.
}
\end{figure*}

The differential cross section for the productions of $ZHH$  
is \cite{Djouadi:1999gv}
\bea
\frac{d \sg [\ee\to Z H H]}{d x_1 d x_2}
=\frac{ \sqrt{2}\, G_F^3 m_Z^6}{384 \pi^3 s}
\frac{v_e^2 + a_e^2}{(1-\mu_Z)^2}
\mathcal{Z}_{HH}(x_1,x_2)
\eea
where $x_{1,2}=2 E_{1,2}/\sqrt{s}$
are the scaled energies of two Higgs bosons, 
$v_e = -1 +4 s_W^2$, $a_e =-1$, and $\mu_i = m_i^2/s$. 
The coefficients of $\mathcal{Z}_{HH}$ in our scenario is
\bea
\mathcal{Z}_{HH} &=&
\left(
\frac{\lm_{HHh}}{y_3-\mu_{hZ}} +
\frac{1}{\mu_Z}
\right)^2 f_0 
+
\frac{1}{2}\left(
\frac{\lm_{HHh}}{y_3-\mu_{hZ}} +
\frac{1}{\mu_Z}
\right)
\frac{f_3}{y_1 + \mu_{HA}}
\\ \nn &&
+
\frac{1}{4 \mu_Z(y_1+\mu_{HA})}
\left[
\frac{f_1}{y_1+\mu_{HA}} + \frac{f_2}{y_2 + \mu_{HA}}
\right] 
\\[5pt] \nn
&&
+
\large\{
y_1 \leftrightarrow y_2
\large\},
\eea
where $y_i=1-x_i$, $x_3 = 2-x_1-x_2$, $\mu_{ij}=\mu_i -\mu_j$, and
the expressions for $f_{0,1,2,3}$ are referred to the Appendix A of Ref.~\cite{Djouadi:1999gv}.

In Fig.~\ref{fig-sig},
we show the total cross section $\sg(\ee\to ZHH)$
as a function of $\tb$ for $M_H=500\gev$, $M_A=400,500\gev$, and $\sqrt{s}=1.5\tev$.
The other parameters vary freely while satisfying the theoretical constraints.
The main outcome in Fig.~\ref{fig-sig}
is that
the theoretical constraints in the form of inequalities
almost determine the total cross section
for the given $M_H$ and $M_A$ if $\tb\gsim 5$.
The cross section is of the order of 10 ab,
implying that probing $\lm_{HHh}$
through the double Higgs-strahlung processes is challenging.
However, the strong correlation between the total cross section 
and the heavy Higgs boson masses shall play an important role in
justifying or invalidating the 2HDM
when we observe a heavy scalar boson.

\section{Conclusions}

In the aligned two Higgs doublet model (2HDM)
with decoupling,
we have studied the theoretical constraints
from the bounded-from-below potential, unitarity, perturbativity, and the vacuum stability.
We found that some parameters $\lm_i$'s in the scalar potential 
become highly enhanced when $M_{H,A,H^\pm}\gg m_h$
and $\tan\beta \gg 1$, 
which signals a breakdown of the perturbative unitariry.
Theoretical constraints in the form of inequalities
play a significant role, which  
allows an extremely narrow parameter space.

When $M_{H} \gsim 500\gev$ and $\tan\beta \gsim 10$,
only a fine line in the parameter space $(\tan\beta,m_{12}^2)$
survives with the given $M_H$, irrespective to the other non-SM Higgs boson masses.
Practically, the inequalities are reduced to an equation,
$m_{12}^2 \tan\beta \simeq C$
where $C$ is a constant with the given $M_H$.
The $m_{12}^2$ is determined by the value of $\tan\beta$ within $\mathcal{O}(1)\gev$.
Moreover the other Higgs boson masses should be similar to the given $M_H$
within $\mathcal{O}(10)\%$.
For the given $M_A$,
the correlation between $\tan\beta$ and $m_{12}^2$
is not as dramatically strong as in the fixed $M_H$ case.
The value of $\tan\beta$ sets $m_{12}$ within $\mathcal{O}(10)\gev$.

We also studied the theoretical constraints
on the triple Higgs boson couplings.
Due to $m_{12}^2 \tan\beta \simeq C$,  
$\lm_{HHh}$ becomes nearly constant 
for $M_{H} \gsim 500\gev$ and $\tan\beta \gsim 10$.
Both $\lm_{HHH}$ and $\lm_{AHH}$ are linearly proportional to $\tan\beta$.
The total cross section of $\ee\to ZHH$ was calculated
for $M_H=500\gev$ at $\sqrt{s}=1.5\tev$.
Because of the almost independence of $\lm_{HHh}$ on $\tan\beta$,
$\sg( \ee\to ZHH ) \sim \mathcal{O}(10)\,$ab is also constant about $\tan\beta$.
The studied characteristics of the aligned 2HDM with decoupling
only from the theoretical constraints
are strong enough to probe or invalidate 
the model in the future collider experiments.

\begin{acknowledgments}
We thank Sin Kyu Kang for useful discussions.
This paper was written as part of Konkuk University's research support program 
for its faculty on sabbatical leave in 2016.
\end{acknowledgments}

\end{document}